\documentstyle[aps,prl,multicol,amsfonts,amssymb,epsfig]{revtex}

\newcommand{\Ref}[1]{Ref.~\cite{#1}}

\newcommand{\be}{\begin{equation}}
\newcommand{\ee}{\end{equation}}
\newcommand{\ba}{\begin{eqnarray}}
\newcommand{\ea}{\end{eqnarray}}                                   
 
\newcommand{\Eq}[1]{Eq.~(\ref{#1})} 

\newcommand{\vev}[1]{\left\langle #1 \right\rangle}

\renewcommand{\>}{\rangle}

\renewcommand{\Im}{{\rm Im}} 
\renewcommand{\Re}{{\rm Re}}

\newcommand{\prt}{\partial} 
\newcommand{\pt}{\partial} 
\newcommand{\sgn}{{\rm sgn}} 
\renewcommand{\t}[1]{{\tilde #1}}

\newcommand{\ie}{{\it ie~}}

\newcommand{\etal}{{\it etal~}} 

\newcommand{\al}{\alpha} 
\newcommand{\bt}{\beta} 
\newcommand{\del}{\delta} 
\newcommand{\Del}{\Delta} 
\newcommand{\eps}{\epsilon} 
 
\newcommand{\ga}{\gamma}

\newcommand{\la}{\lambda} 
 
\newcommand{\om}{\omega}

\newcommand{\Th}{\Theta} 
\newcommand{\si}{\sigma}

\newcommand{\cH}{ {\cal H} }

\newcommand{\cL}{ {\cal L} }

\def\journal #1, #2, #3, 1#4#5#6{{\sl #1~}{\bf #2}, #3 (1#4#5#6) }

\def\prb{\journal Phys. Rev. B, }

\def\np{\journal Nucl. Phys., }

\renewcommand{\v}[1]{{\bbox #1}}

\begin{document}
\draft
\widetext

\title{
Non-equilibrium 2-Channel Kondo model for quantum dots
}

\author{Xiao-Gang Wen}
\address{
Department of Physics,
Massachusetts Institute of Technology,
Cambridge, MA 02139, USA
}

\date{Sept, 1998}
\maketitle

\widetext
\begin{abstract}
\rightskip 54.8pt
We find that, under certain condition, a quantum dot with odd number of
electron and coupled to two leads  can be described by a non-equilibrium
2-channel Kondo model,
when the two leads has a large voltage bias between them.
The model is exactly soluble and can be mapped into a free fermion system,
even in the presence of external magnetic field and other relevant
perturbations.
All (dynamical) correlation functions can be calculated.  The fixed point of
the 2-channel Kondo model is a non-equilibrium fixed point and is different
from the usual 1-channel and 2-channel Kondo fixed point.
\end{abstract}


\begin{multicols}{2}

\narrowtext

In a recent experiment, Goldhaber-Gordon \etal\cite{Gold} and Cronenwett
\etal\cite{Leo} observed a Kondo effect in quantum dot systems predicted by
theories.\cite{Ng,Gla,Kaw,Mei,Hers,Lev,Win,Wan} The Kondo effect in quantum
dot systems is very interesting since many parameters can be adjusted, which
allow us to probe many different regime/aspects of the Kondo effects.

In this paper we are going to study the Kondo effect in quantum dot
systems in the limit where the charge fluctuations on the dot
can be ignored, and when there is a large voltage bias across the two leads.
Under certain conditions as will be indicated bellow, the properties of the
system is  described by a 2-channel Kondo model, with non-equilibrium effect
included as a perturbation.  This 2-channel Kondo model
can be solved exactly using a Majorana fermion approach even in the presence
of some relevant perturbations, such as finite temperatures,
finite magnetic fields, non-equilibrium tunneling between leads,
and finite changes in the relative strength of the
coupling between the two leads and the dot.  Many properties of the system
(including dynamical properties) can be calculated exactly.  The fixed point
of the 2-channel Kondo model studied here is a non-equilibrium fixed point
which is different from the usual 2-channel Kondo fixed point. For example the
dot-spin correlation behaves as $t^{-2}$ here instead of $t^{-1}$ for the
usual 2-channel Kondo  fixed point.

We start with the following model Lagrangian for the quantum dot
\begin{eqnarray}
\cL &=& i \psi^\dag_{i\al} (\prt_t+v \prt_x
+ i \frac{1}{2}eV \tau^3_{ij})\psi_{j\al}
- \v h \cdot \v S \del(x)
\nonumber\\
&& -2 \la \del(x) \v S \cdot (u^* \psi^\dag_{1\al} + v^*  \psi^\dag_{2\al})
\frac{\v \si_{\al\bt}}{2}  (u \psi_{1\bt} + v  \psi_{2\bt})
\label{Lori}
\end{eqnarray}
where $V$ is the voltage difference between the first and the second lead, and
$\v h$ is the coupling of external magnetic field to the dot-spin.  Note that
the spin-up and the spin-down Fermi surfaces in the leads remains to have the
same energy even in the presence of external magnetic field, thus there is no
coupling between the external magnetic field and the lead-spins in our model.
$\la u$ and $\la v$ describe the strength of coupling between the dot and the
two leads.

After redefine the $\psi$ field in \Eq{Lori}:
$\psi_{i\al} \to e^{-i  \frac{1}{2}eV \tau^3_{ij} t } \psi_{j\al}$,
the above  Lagrangian can be rewritten as
\begin{eqnarray}
&&\cL = \cL_{Knd} +\cL_t +\cL_h \label{L} \\
&&\cL_{Knd} = i \psi^\dag_{i\al} (\prt_t+v \prt_x)\psi_{j\al} \nonumber\\
&&\ \
-2 \la \del(x) \v S \cdot
\left( |u|^2 \psi^\dag_{1\al} \frac{\v \si_{\al\bt}}{2}  \psi_{1\bt}
+ |v|^2 \psi^\dag_{2\al} \frac{\v \si_{\al\bt}}{2}  \psi_{2\bt} \right)
\label{L_K}  \\
&&\cL_t = -  2 \la_t \del(x) \v S \cdot
\left( e^{ieV t}
u^* v \psi^\dag_{1\al} \frac{\v \si_{\al\bt}}{2}  \psi_{2\bt}
+ h.c. \right)
\label{Lt}
\\
&&\cL_h = - \v h \cdot \v S \del(x)
\end{eqnarray}
where $\la_t=\la$. We see that the finite bias is described by a time 
dependent term after the mapping. 
$\la_t$ is the amplitude of the spin-flip tunneling between the two leads, and
$\la$ is the spin-exchange coupling between the dot-spin and the lead-spins.
\begin{figure}
\epsfxsize=1.5truein
\centerline{\epsffile{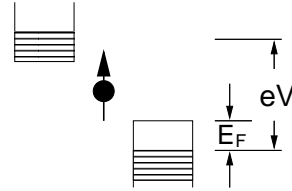}}
\caption{
The tunneling can be reduced by a change in the density of state.
}
\label{fig1}
\end{figure}

The above system is a 1-channel Kondo system when $eV=0$ since $\la_t=\la$.
However, when $eV$ is large, the spin-exchange term and the spin-flip
tunneling term become very
different. If one can make $\la_t\ll \la$ in the large $eV$ limit, then the
system becomes a 2-channel Kondo problem. One way to reach this limit
is to use
the energy dependence of the density of states. For example, one may use
n-type semiconductor as one lead and p-type as the other lead (\ie put the
quantum dot in the depletion layer of a diode). When $eV>E_F$,
we have $\la_t\ll \la$ since the tunneling is blocked (see Fig. \ref{fig1}).
Certainly, to see the Kondo effect, we need $E_F$ to be less then the level 
spacing in the dot.

Even when $\la = \la_t$ at high energies, we can effectively have $\la>\la_t$
at low energies. This is because at energy scales above $eV$, the system is a
1-channel Kondo system, and $\la$ and $\la_t$ flow together to larger values
as we lower the energy scale.  When energy scales are less then $eV$, $\la_t$
can no longer flow.  If the value of $\la_t$ is small enough at $eV$ (\ie if
$eV \gg T_K$ where $T_K$ is the 1-channel Kondo temperature for $eV=0$), the
system behave like a 2-channel Kondo system and $\la$ can continue to flow to
even larger values as we lower the energy scales further.  Thus at low
energies, we effectively have $\la_t > \la$.  However, since $\la$ and $\la_t$
flow as $\ln(E)$, it may be difficult to achieve $\la \gg \la_t$ at low
energies.

When $\la_t \ll \la$ (at low energies), we may first set $\la_t=0$, and study
the 2-channel Kondo problem described by $\cL_{Knd}$.  The 2-channel Kondo
system $\cL_{Knd}$ has a 2-channel Kondo fixed point, which can be reached by
tuning the relative strengths of the couplings between the dot and the two
leads to $|u|^2=|v|^2$, and lowering the temperature below a 2-channel Kondo
temperature $T_{2K}$.  Then, we can include back the $\cL_t$ term to study its
effect on the 2-channel Kondo fixed point.
As we will see later that, at the fixed point,
the effects of the $\cL_t$ term (as well as many other
terms, such as the coupling of dot-spin to the external magnetic field)
can be calculated exactly.

In the following we are going to introduce
a Majorana fermion approach (which is
a combination of the current algebra approach in \Ref{A,AL} and the Majorana
fermions approaches in \Ref{EK,LMaj,Ye}) to solve the 2-channel Kondo fixed
point.
The Majorana fermion approach used here has a full spin rotation symmetry,
and is similar to the one used in \Ref{CIT}.
The 2-channel Kondo Hamiltonian is given by
\begin{equation}
  H=H_f+H_K=
iv \int dx\ \left( \psi^\dag_{i\al}\prt_x \psi_{i\al} +
\la  \v J_s(0) \cdot \v S \right)
\end{equation}
where $H_f$ is the free fermion Hamiltonian.
The above 2-channel Kondo system flows to a fixed point with coupling
$\la=\la_0\equiv \pi v$.
In the rest of the paper we are going to only consider physics
at energy scale much lower then the Kondo temperature $T_{2K}$
and concentrate on the fixed point
Hamiltonian with $\la=\la_0$.

Introducing the charge $U_{\rm c}(1)$, spin $SU_{\rm s}(2)$, and flavor
$SU_{\rm ch}(2)$ densities
(we will call the $SU_{\rm ch}(2)$ quantum numbers as flavors):
\begin{equation}
    J = \psi^\dag_{i\al}\psi_{i\al} ,\ \
 \v J_s = \psi^\dag_{i\al}\frac{\v \si_{\al\bt}}{2}\psi_{i\bt}, \ \
  J_f^A = \psi^\dag_{i\al}\frac{\tau^A_{ij}}{2}\psi_{j\al}.
\end{equation}
we can rewrite the 2-channel Kondo Hamiltonian as
\begin{eqnarray}
 H_f &=& \frac{\pi v}{4}J^2 + \frac{\pi v}{2}\v J_s^2
+ \frac{\pi v}{2}\v (J_f^A)^2  \nonumber\\
H_K &=&  \la \del(x) \v S \cdot \v J_s
\end{eqnarray}
Introducing $\t {\v J}_s(x)=\v J_s(x)+ \del(x) \v S$
which satisfies the same algebra as
$\v J_s$,\cite{A} we can rewrite the above Hamiltonian as
\begin{equation}
 H_{\rm 2CK}
=\frac{\pi v}{4}J^2 + \frac{\pi v}{2}{\t{\v J}_s}^2 + \frac{\pi v}{2} (J_f^A)^2
+{\rm const.}
\label{Hfix}
\end{equation}
if $\la =\la_0$. In terms of ${\t{\v J}_s}$, the 2-channel Kondo Hamiltonian at
the fixed point $\la=\la_0$ takes the same form as a free fermion Hamiltonian.

In the following, we are going to use Majorana fermions to describe our
2-channel Kondo system.
Introduce eight Majorana fermions $\sqrt 2 \Re \psi_c$, $\sqrt 2 \Im \psi_c$,
$\eta_f^A|_{A=1,2,3}$
and $\eta_s^a|_{a=1,2,3}$,\cite{LMaj}
where $\psi_c$ carries $\sqrt 2$ unit of the
$U_{\rm c}(1)$ charges, $\eta_f^A|_{A=1,2,3}$ form a flavor triplet and
$\eta_s^a|_{a=1,2,3}$ a spin triplet. The free Hamiltonian $H_f$ 
can be rewritten as
\begin{equation}
H_f = iv \int dx\ \left( \psi_c^\dag\prt_x \psi_c
+ \frac12 \eta_f^A \prt_x \eta_f^A + \frac12 \eta_s^a \prt_x \eta_s^a \right)
\end{equation}
The free theory is described by the
 $SU_{\rm s}(2)_2\times SU_{\rm ch}(2)_2\times U_{\rm c}(1)$
KM algebra.
In terms of the Majorana fermions, the currents $J_c$, $\v J_s$, and $J_f^A$
take the form
$ J_c = \sqrt 2 \psi_c^\dag\psi_c $,
$ J_s^a =  \eps^{abc}\eta_s^a\eta_s^b/2 $, and
$ J_f^A =  \eps^{ABC}\eta_f^A\eta_f^B/2$.

Let us concentrate on the spin part described by
\begin{equation}
 H_s= iv \int dx\ \frac12 \eta_s^a \prt_x \eta_s^a
\end{equation}
Assume the system is finite $-L/2 < x < L/2$, and $\eta_s^a$ satisfy the
anti-periodic boundary condition: $\eta_s^a(x)=-\eta_s^a(x+L)$. In this case
the ground state $|0\>$ is a spin singlet. The total Hilbert space contains
two sectors: $\cH^s=\cH^s_0\oplus \cH^s_1$.
$\cH^s_0$ containing even number of fermions
is the space generated from $|0\>$
by applying any number of the $\v J_s$ operators,
while $\cH^s_1$ containing odd number of fermions is generated
by applying any number of the $\v J_s$ operators and one $\eta_s^a$ operators.

Now let us add the Kondo term $H_K$ at the critical coupling $\la_0$.
The spin part is still described by the $SU_{\rm s}(2)_2$ KM algebra
in terms of the new $SU_{\rm s}(2)$ current $\t {\v J}_s$.
Notice that $H_K$
conserves the even-odd fermion numbers.
Thus the total Hilbert space still contains two sectors:
$\cH^s=\cH^s_{1/2}\oplus \cH^{s\prime}_{1/2}$, where $\cH^s_{1/2}$ (or
$\cH^{s\prime}_{1/2}$) carries even-number (or odd-number) of fermions.
The system can be solved using KM algebra. 
The ground states in each sector carry spin 1/2 due to the
added dot-spin. All the states in each sector are generated by
applying $\t {\v J}_s$'s on the corresponding ground states and the states
in $\cH^s_{1/2}$ and $\cH^{s\prime}_{1/2}$ have a one-to-one correspondence.
We see that the Kondo coupling induces the following simple changes in the
Hilbert spaces:\cite{AL}
\begin{equation}
\cH^s_0\to \cH^s_{1/2},\ \ \ \cH^s_1\to \cH^{s\prime}_{1/2}
\label{cHcH}
\end{equation}
We can introduce new Majorana fermions to represent the new
$SU_{\rm s}(2)$ current:
$
 \t J_s^a = \frac{1}{2} \eps^{abc}\t \eta_s^a\t \eta_s^b .
$
$\t \eta_s^a$ carry odd fermion numbers and map between even- and odd-fermion
sector. When $x\neq 0$, $\t \eta_s^a(x)=\sgn(x)\eta_s^a(x)$.

Note that the total Hilbert space at the fixed point has a structure
$\cH^s=\cH^s_{1/2}\otimes \{|0\>, |1\>\}$. We can introduce a fermion operator
$d^\dag$ that map the
even-fermion state $|0\>$ to the odd-fermion state $|1\>$:
$|1\>=d^\dag |0\>$. Note that $d$ and $d^\dag$ commute with $\t {\v J}_s$,
and do not appear in the fixed point Hamiltonian (which contains only
$\t {\v J}_s$). Thus $d$ is a free fermion operator which carries no energy.
Using the commutation relation (which actually defines a spin triplet
primary field)
\begin{equation}
 [\t J_s^a(x), S^b]=i\eps^{abc}\del(x) S^c
\end{equation}
and the fact that $\v S$ carry even-fermion numbers, we find that
the dot-spin operator $\v S$ is given by
\begin{equation}
 \v S = \frac{\sqrt a}{2} \hat b  \t {\v \eta}_s(0)
\end{equation}
where $\hat b =d+d^\dag$.
The normalization coefficient is obtained by noticing that
$\{ \t \eta^a_s(0),  \t \eta^b_s(0) \}=\del^{ab} \del(0)$ and $\del(0)=1/a$
if we choose a finite short distance cut-off $a$.
With the fixed point Hamiltonian
\begin{equation}
 H_s+H_K= iv \int dx\ \frac12 \t \eta_s^a \prt_x \t \eta_s^a
\end{equation}
we can now easily calculate the $\v S$ correlation.

Now let us add back the non-equilibrium
tunneling term $\cL_t$ and include the channel asymmetry term
\begin{equation}
\v S \cdot
\psi^\dag_{i\al}(0) \tau^3_{ij} \v \si_{\al\bt}  \psi_{j\bt}(0)
\label{asym}
\end{equation}
Both terms are spin singlet and flavor triplet. Thus the leading contribution
to those operators at the 2-channel Kondo fixed point
comes form the flavor triplet primary fields which have dimension 1/2.
To obtain the Majorana fermion representation of the flavor
triplet primary fields,
let us analyze the Hilbert space of the spin and the flavor
sectors (with zero-charge).
Without the Kondo coupling, the Hilbert space has a form
$\cH^{sf}=\cH^s_0\otimes\cH^f_0 \oplus \cH^s_1\otimes\cH^f_1 $, where
$\cH^s_{0,1}$
are defined above for the spin sector, and $\cH^f_{0,1}$ are defined similarly
for the flavor sector.
The sector $\cH^s_0\otimes\cH^f_0$ (or $\cH^s_1\otimes\cH^f_1$) is
generated by $J^a_s$ and $J_f^A$ from $|0\>$ (or
$\eta_s^a\eta_f^A|0\> \sim \psi^\dag_{i\al}
\si^a_{\al\bt}\tau^A_{ij} \psi_{j\bt}|0\>$).
After we include the Kondo coupling, according to
\Eq{cHcH}, the Hilbert space becomes
$\cH^{sf}=\cH^s_{1/2}\otimes\cH^f_0 \oplus \cH^{s\prime}_{1/2}\otimes\cH^f_1 $.
The above structure of the Hilbert space tells us that the spin triplet
Majorana
fermion operator $\t \eta^a_s$ are unphysical since they map a state in
$\cH^s_{1/2}\otimes\cH^f_0$ to a state in $\cH^{s\prime}_{1/2}\otimes\cH^f_0 $
which is outside of our physical
Hilbert space. Similarly the flavor triplet Majorana
fermion operator $\eta^A_f$ are also unphysical. The only physical
spin triplet primary fields are $\v S$ and the only physical
flavor triplet primary fields are $\hat b \eta_f^A$.
We note that the flavor triplet primary fields are quadratic in the fermion
operators. Our model for the fixed point remains to be a free fermion theory
even after we include
the relevant perturbations represented by $\hat b \eta_f^A$.
This allows us to study exactly
the effect of $\cL_t$ term and the channel asymmetry term
on the 2-channel Kondo fixed point.

More generally, we find the following exactly soluble model
\begin{eqnarray}
 H &=& H_{\rm 2CK} + H_a + H^\prime_a
\nonumber\\ &&\ \
+ H_t +H^\prime_t + H_{h}+H^\prime_{h}
+H^\prime_{fs}
\\
H_{\rm 2CK} &=&
-iv \int dx\ \left( \psi_c^\dag\prt_x \psi_c
+ \eta_f^A \frac{\prt_x}{2} \eta_f^A
+ \t \eta_s^a \frac{\prt_x}{2} \t \eta_s^a \right)
\\
H_a &=& i\ga_a \hat b \eta_f^3(0)
\label{Ha}
\\
H^\prime_{a} &=& \ga^\prime_{a} i \eps^{3AB} \eta_f^A(0)\eta_f^B(0)
\\
H_t &=& i\frac{1}{\sqrt 2} \ga_{t} e^{ieVt} \hat b \psi_f^\dag(0) +h.c.
\\
H^\prime_{t} &=& i\frac{1}{\sqrt 2} e^{ieVt}\psi_f^\dag(0) \left(
 \ga^\prime_{tf} \eta_f^3(0) + \ga^\prime_{ts} \eta_s^3(0) \right)  +h.c.
\\
H_{h} &=&
i \hat b \v \ga_s \cdot \t {\v \eta}_s(0)
\\
H^\prime_{h} &=&
 \ga_s^{\prime a} i \eps^{abc} \eta_s^b(0)\eta_s^c(0)
\\
H^\prime_{fs} &=&
 i \ga_{fs}^{aA} \eta_f^A(0)\eta_s^a(0)
\end{eqnarray}
where $\psi_f=(\eta_f^1-i \eta_f^2)/\sqrt 2$
and $\v \ga_s=  \frac{\sqrt a}{2} \v h $.
This Hamiltonian describes what we call the non-equilibrium
2-channel Kondo model.  $H_{\rm 2CK}$ in $H$
describes the 2-channel Kondo fixed point.
All other terms represent relevant or marginal perturbations around the
2-channel Kondo fixed point.
$H_a$ and $H^\prime_a$ are induced by the asymmetry in the coupling between the
channel spins and the dot-spin. $H_t$ and $H^\prime_{t}$ are the
non-equilibrium tunneling
terms caused by the finite voltage bias between the two channels. $H_t$
corresponds to the spin-flip tunneling described by
$\cL_t$ and $H^\prime_{t}$ corresponds
to the direct tunneling described by $\psi_{1\al}^\dag\psi_{2\al} + h.c.$
and $\psi_{1\al}^\dag\si^3_{\al\bt}\psi_{2\bt} + h.c.$
$H_{h}$ is the coupling between a magnetic field $\v h$ and the dot-spin
$\v S$.

To calculate the physical properties of the above system
we also need to specify the ``boundary''
condition, \ie how incoming fermions are distributed.
The boundary condition here is that all income branches contain no
Majorana fermions, \ie all $k>0$ levels are empty.

Next we simplify the Hamiltonian using
the transformation $\psi_f(x,t)\to e^{ieV(t+v^{-1} x) }\psi_f(x,t)$
which changes the $eV$ in $H$ to zero.
However, the transformation also changes the boundary condition
for the incoming branches:
$\psi_f$ is now filled up to energy $eV$, but the
$k>0$ levels for all other Majorana fermions remain to be empty.

For simplicity, let us drop the sub-leading terms $H^\prime$'s.
Among the eight Majorana fermions in $H$ (now with $eV=0$)
only one linear combination couples to
$\hat b$:
\begin{equation}
H_b =
\int dx\ (-iv)\eta_b \frac{\prt_x}{2} \eta_b  +i \ga \hat b \eta_b(0)
\end{equation}
where
\begin{eqnarray}
\eta_b &=& \ga^{-1}[\ga_a \eta^3_f + \Re (\ga_t) \eta_f^1
+ \Im (\ga_t) \eta_f^2 + \v \ga_s\t {\v \eta}_s ]
\\
\ga &=& \sqrt{ \ga_a^2 + |\ga_t|^2 + \v \ga_s^2}
\end{eqnarray}
The equation of motion are\cite{CFW}
\begin{eqnarray}
\pt_t \eta_b(x,t) &=& -v \pt_x \eta_b(x,t) - \ga \hat b(t)\del(x)
\\
 \pt_t \hat b(t)& = & -2 \ga \eta_b(0,t)
 = -\ga(\eta_b(0^+,t)+ \eta_b(0^-,t))
\end{eqnarray}
If we expand
$ \eta(x) = \sum_k \eta_- (k) e^{ikx}$ for $x<0$ and
$ \eta(x) = \sum_k \eta_+ (k) e^{ikx}$ for $x>0$,
then the scattering is simply an energy dependent phase shift:
$ \eta_+ (k) =e^{i\phi_k} \eta_- (k) $
where
\begin{equation}
 e^{i\phi_k}=\frac{iv^2k-\ga^2}{iv^2k +\ga^2}
\label{phik}
\end{equation}
This allows us to determine how the eight Majorana fermions,
$(\eta_1,...,\eta_8)=(\sqrt 2 \Re \psi_c, \sqrt 2 \Im \psi_c, \eta_f^A, \t
\eta_s^a)$, scatter across the
$d$ level at $x=0$:
\begin{equation}
 \eta_{I+}(k)= M_{IJ} \eta_{J-}(k),\ \
M=1+(e^{i\phi_k}-1) \frac{\ga_I \ga_J}{\ga^2}
\end{equation}
where
$(\ga_1,...,\ga_8)=(0,0,\Re(\ga_t),\Im(\ga_t),\ga_a,\ga_s^1,\ga_s^2,\ga_s^3)$,
and $\eta_{I\pm}(k)$ are the Fourier components of $\eta_I(x)$ for $x>0$
and $x<0$. The large phase shift $\phi_k=\pi$ at $k=0$ indicates a resonance
at the Fermi level.
This setup allows us to easily calculate the total tunneling current\cite{CFW}
$I_t=ev \int \frac{dk}{2\pi} \vev{ \psi^\dag_{f+}(k) \psi_{f+}(k)
- \psi^\dag_{f-}(k) \psi_{f-}(k)}$. We find that
\begin{eqnarray}
I_t &=& e \int \frac{d k}{2\pi} \frac{ v |\ga_t|^2  }{
(|\ga_t|^2 +\ga_a^2 + \v \ga_s^2)^2  +  v^4 k^2 } \times
\nonumber\\
&&\ \
 \Big[ (|\ga_t|^2+\v \ga_s^2+\ga_a^2)(n(vk+eV)-n(vk-eV)) \Big]
\label{It}
\end{eqnarray}
where $n(\eps)=1/(e^{\eps/T}+1)$.
Unfortunately $I_t$ having a very smooth dependence on $V$ cannot reveal
the resonance at the Fermi surface when $eV$ is large.
One need to use the noise in $I_t$ near $\om =eV$ to probe the
resonance.\cite{Anna} 

The effect of $H^\prime$ is to generate additional {\em energy independent}
rotations among $\eta_I$'s. It changes $M$ to
$M=1+(e^{i\phi_k}-1) \ga^+_I \ga^-_J$, where $\ga^\pm$ are two unit vectors.

The equation of motion for $\hat b$ also implies that
$
 \hat b=  -i \int \frac{dk}{2\pi} \frac{\ga}{vk}(1+e^{i\phi_k} )\eta_-(k)
$
which satisfies $\hat b^2=1$.
This expression allows us to calculate the response function
$iD_f(t)=\Th(t) \vev{[F^3(t), F^3(0)]}$ where $F^3$ is the channel asymmetry
operator $F^3=i\hat b \eta^3_f$ (which comes from \Eq{asym}).
We find that (assuming $\ga_a=\v \ga_s=0$ and $eV\to \infty$ for simplicity)
\begin{eqnarray}
 \Im D_f(\om) &=&  \frac{1}{2\pi v} \int d\nu\
 \frac{\ga^2 v^{-1}}{\nu^2 + \ga^4v^{-2}} ( n(\nu-\om)-n(\om-\nu) )
\nonumber\\
 D_f(\om) &=& -\frac{1}{\pi}\int d\nu\ \frac{\Im D_f(\nu)}{\om-\nu+i0^+}
\end{eqnarray}

Assume $\ga_a$ depends on a gate voltage $V_g$ which control the strength
of coupling between the dot and the leads.
Changing $V_g$ will generate
a term $\Del V_g \frac{d\ga_a}{dV_g} F^3$ in Hamiltonian. Thus
$Q=\frac{d\ga_a}{dV_g} F^3$ is the charge operator that couples to $V_g$.
The induced charge is
\begin{equation}
 Q(\om)=\left(\frac{d\ga_a}{dV_g}\right)^2 D_f(\om) \Del V_g(\om)
\end{equation}
We see that $C=\left(\frac{d\ga_a}{dV_g}\right)^2\Re D_f(\om)$
is the capacitance,
and $\si =\om \left(\frac{d\ga_a}{dV_g}\right)^2\Im D_f(\om)$
is the conductance in parallel with the capacitance.
We find that, for $\om> \ga^2v^{-1}$, $C\sim \ln (T_{2K}/\om)$
and $\si\propto \om$,
while for $\om < \ga^2v^{-1}$, $C\sim \ln( vT_{2K}/\ga^2)$ and $\si\propto
\om^2$. Measuring $C$ and $\si$ by applying an AC component in the
gate voltage $V_g$ may allow us to
probe the resonance associated with the fixed point. The behavior for $\om>
\ga^2v^{-1}$ is the behavior for usual 2-channel Kondo fixed point. 

Similarly we can also calculate the dot-spin correlation exactly.  In the
presence of the $\cL_t$ term, the correlation for dot-spin has the usual
2-channel Kondo behavior $1/t$ for $t< v/\ga^2$, and crosses over to $1/t^2$
behavior for $t> v/\ga^2$.

We would like to repeat that all of the above results are valid only when
$\ga^2/v $ and $T$ are much less than the Kondo temperature $ T_{2K}$.

XGW is supported by NSF Grant No. DMR--97--14198 and by NSF-MRSEC Grant
No. DMR--94--00334.


\end{multicols}

\end{document}